\documentclass[aps,prl,twocolumn,groupedaddress]{revtex4}

\usepackage{graphicx}

\begin{document}


\title{Understanding self-assembled nanosphere patterns}

\author{Ferenc J\'arai-Szab\'o}
\author{Simion A\c stilean}
\author{Zolt\'an N\'eda}
\email[]{zneda@phys.ubbcluj.ro}

\affiliation{Babe\c s-Bolyai University, Dept. of Physics, str.
Kog\u alniceanu 1, RO -- 400084, Cluj-Napoca, Romania}

\date{\today}

\begin{abstract}
Patterns generated by a colloidal suspension of nanospheres drying
on a frictional substrate are studied by experiments and computer
simulations. The obtained two-dimensional self-assembled
structures are commonly used for nanosphere lithography.  A
spring-block stick-slip model is introduced for simulating the
phenomenon and the influence of several controllable parameters on
the final structure is investigated.  The model successfully
reproduces the experimentally observed patterns and the dynamics
leading to pattern formation is revealed.
\end{abstract}


\maketitle

\begin{figure}
 \includegraphics[width=8.5cm]{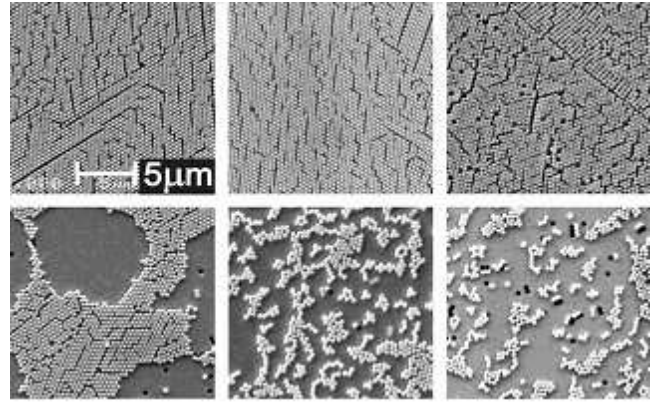}%
 \caption{Characteristic two-dimensional patterns obtained from a colloidal
          suspension of 220 nm diameter nanospheres drying on a silica
          substrate.\label{fig1}}
 \end{figure}

The so-called bottom-up approach for the fabrication of
nanostructures starting from stable building blocks such as
molecules or nanoparticles has become an increasingly popular
topic in nanoscience and nanotechnology. Thanks to the efforts of
nanochemists, during the past decades various nanoparticles of
polystyrene, silica, noble-metal and semiconductor, nearly
monodisperse in terms of size, shape, internal structure, and
surface chemistry, can be produced through a reliable, standard
manufacturing process. Using these nanoparticles as building
blocks, the synthesis of long-range-ordered monolayers and films
of colloidal nanocrystals has been in particular focus. The
revolutionary development of photonic crystals triggered efforts
to get innovative methods for crystallizing polystyrene colloids
and creating new crystal structures \cite{xia2000,wang2004}. The
use of two-dimensional (2D) self-assembled array of
nanometer-sized polystyrene spheres as deposition mask is known as
NanoSphere Lithography (NSL) \cite{haynes2001}. The homogeneous
arrays of nanoparticles produced using NSL are potentially useful
in studies of size-dependent optical, magnetic, catalytic and
electrical transport properties of materials
\cite{haynes2001,rybczynski2003,kempa2003,murray2004}. NSL is now
recognized as a powerful fabrication technique to inexpensively
produce nanoparticle arrays with controlled shape, size and
interparticle spacing. A fundamental goal for further progress in
NSL is the development of experimental protocols to control the
interactions, and thereby the ordering of nanoparticles on solid
substrates \cite{chabanov2004,aizenberg2000}. However, it is more
and more clear that due to the rich physics and chemistry
underlying the formation of nanoparticle arrays from colloidal
suspensions, the likelihood of structures other than close-packed
networks forming during solvent evaporation is very high
\cite{pileni2001,ge2000}. Therefore a major motivation for
theoretical research in this field remains the challenge to
understand how ordered or complex structures form spontaneously by
self-assembly, and how such processes can be controlled in order
to prepare structures with a pre-determined geometry
\cite{vasco2004}. The present study intends to contribute in this
sense by proposing a model that can be easily studied through
computer simulations and it is able to qualitatively reproduce the
wide variety of observed patterns. We focus on an experimentally
simple case, when 2D self-assembled arrays of nanometer-sized
polystyrene spheres will form from a colloidal suspension which is
drying on a substrate. Some characteristic patterns obtained as a
result of this phenomenon are present in Figure~\ref{fig1}.

The model is rather similar with the spring-block stick-slip model
successfully used for describing fragmentation structures obtained
in drying granular materials in contact with a frictional
substrate \cite{leung2000,neda2002}. The new feature of the
present model is that a predefined lattice is not considered
anymore. Proceeding in such manner the final lattice structure
appears as a result of the underlying interactions governing the
dynamics of the system and the existing geometrical constraints.

\begin{figure}
 \includegraphics[width=8.5cm]{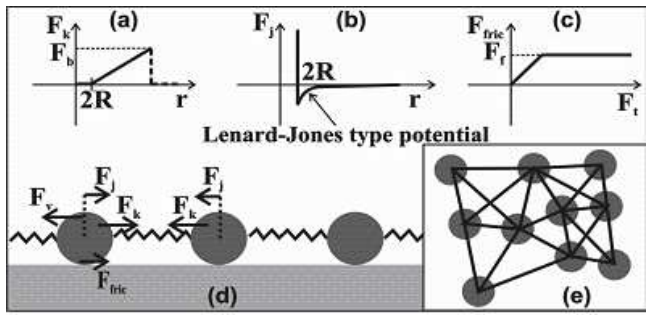}%
 \caption{Sketch of the spring-block stick-slip model. Figure~(a) and (b) shows
          the distance dependence of the tension $F_k$ in the spring and the
          Van-der Waals type force $F_j$ between nanospheres, respectively.
          Figure~(c) illustrates the friction force $F_{fric}$ between nanospheres
          and substrate, as a function of total force $F_t$ acting on nanospheres.
          Figure~(d) presents the main elements of the model, and Figure~(e) is an
          example for a simplified spring-block monolayer system.\label{fig2}}
 \end{figure}

The model is two-dimensional; its main elements are blocks which
can move on a frictional substrate and springs connecting these
blocks (Figure~\ref{fig2}d). Disks, all with the same radius $R$,
model nanospheres, while springs which are characterized by their
variable length and similar spring constants $k$, represents water
bridges between them. The length of a spring is defined as the
distance between perimeters of connected disks. There is also a
Lenard-Jones type interaction-force $F_j$, between each pair of
disk. This is characterized by a strong, almost hard-core type
repulsion which forbids disks to interpenetrate each other and by
a weak attractive type force, accounting for the electric Van der
Waals type interaction between nanospheres (Figure~\ref{fig2}b).
The friction between disks and surface equilibrates a net force
less than $F_f$ (Figure~\ref{fig2}a). Whenever the total force
acting on a disk exceeds $F_f$, the disk slips with an over-damped
motion. The tension in each spring is proportional with the length
of the spring ($F_k = k \cdot l$), and has a breaking threshold
$F_b$ (Figure~\ref{fig2}a).

Initially disks are randomly distributed and connected by a
network of springs (Figure~\ref{fig2}e). We put springs between
those spheres, for which their center can be connected without
intersecting another sphere (this condition will be referred later
as the geometric condition). An initially pre-stressed
spring-block network is thus constructed. During each simulation
step the spring constant is fixed and the system relaxes to an
equilibrium configuration where the tension in each existing
spring is lower than the breaking threshold $F_b$, and the total
net force acting on each disk is lower in magnitude than the
slipping threshold $F_f$. This relaxation is realized through
several relaxation steps. The time length dt for each relaxation
step is taken as unity ($dt = 1$). (Since we are interested mostly
in the final, stable structure, connection with real time is not
crucial for us, and we can define time units in an arbitrary
manner.) Considering a classical molecular dynamics simulation for
relaxation would be very time-consuming. Following the method used
for simulating drying processes in granular media
\cite{leung2000,neda2002}, we choose thus a simplified dynamics,
where the connection with real time is lost, but the relaxation
remains realistic:

(1) For all springs the tension $\left| \vec{F}_k^{ij} \right|$ is
compared with $F_b$. If $\left| \vec{F}_k^{ij} \right| > F_b$, the
spring is broken and taken away from the system.

(2) The total forces $\vec{F}_t^{i} = \sum_p{\left(
d_{ip}\vec{F}_k^{ip} + \vec{F}_j^{ip} \right)}$ acting on disks
are calculated (the sum is over all the other disks $p$, $d_{ip}$
is 1 if the disks are connected by a spring and 0 otherwise, the
subscripts $k$ and $j$ denotes elastic forces from springs and Van
der Waals type forces between disks, respectively).

(3) Each disk is analyzed. If the magnitude of the total force
$\left| \vec{F}_t^{i} \right|$ acting on a disk is bigger than
$F_f$, then the disk will slip with an over-damped motion governed
by viscosity $\eta$, and its position will be changed by:
$\vec{dr}^i = \vec{F}_t^i dt / \eta$. The repulsive part of the
Lenard-Jones potential forbids the spheres to slide on each other
and the presence of viscosity eliminates unrealistic oscillations.

(4) During the motion of a disk it can happen that another spring
is intersected. This intersected spring will brake and it will be
taken away from the system.

(5) After all disks have been visited and their possible motions
done, the springs that fulfill the considered geometrical
condition and for which the tension is lower than the breaking
threshold are redone. By this the rearrangement of water between
nanospheres is modelled.

This concludes one relaxation step. The relaxation is continued
until a relaxation step is finished without having any spring
breaking or disk slipping event. Since in our algorithm it takes a
very long time to achieve a perfect relaxation, we introduce a
tolerance, and assume the relaxation completed when the maximal
slip is smaller than this tolerance. After the relaxation is done,
we proceed to the next simulation step and increase all spring
constants by an amount $dk$. This models that water bridges are
getting thinner and the capillary forces increase the tension in
them as the water evaporates. The system is relaxed for the new
spring-constant value, and the spring constant is increased again,
until all springs are broken or a stable limiting configuration is
reached.

The above dynamics can be easily implemented on computer and
relatively big systems with over 10000 of disks can be simulated
in reasonable computational time. Several types of boundary
conditions can be considered. One possibility is the use of free
boundary condition which can be realized in a simple manner by
positioning initially the disks inside a circle to minimize the
effect of edges.  Another possibility is to consider fixed
boundary conditions. This can be realized for example by
positioning again the disks inside a circle and considering a
chain of fixed disks on the chosen circle. These fixed disks are
then connected with geometrically allowed springs to other disks.
One can also consider periodic boundary condition and position
initially the disks inside a rectangle.  It is quite easy to check
that the considered boundary condition will influence the final
stable structure only in the vicinity of boundaries. In the bulk,
far from the boundaries, the obtained structures are rather
independent of the imposed boundary conditions.  Simulation
results presented here are taken from a small central part of
systems with $N > 10000$ disks using fixed boundary conditions.

The model, as described above, has several parameters: the value
of the static friction $F_f$, the spring breaking threshold value
$F_b$, the initial value of spring constants $k_{ini}$, the spring
constant increasing step $dk$, the viscosity $\eta$, the
parameters of the Lenard-Jones potential, the radius of disks $R$,
and the initial density of nanospheres $\varrho = S / \left( N \pi
R^2 \right)$ (where $S$ is the simulation area). By direct
simulations one can verify that if small enough $dk$ is chosen, in
the quasi-static limit the final structures are not
sensitively influenced by this parameter. The model will only work
for viscosity values chosen between reasonable limits, and for
these viscosity values the final patterns are rather similar.
(Choosing a too small viscosity will result in unrealistic
oscillations of disks, while a too high value will make the disk
slip too small and increase considerably the simulation time). It
is desirable to choose the value of $k_{ini}$ small, to simulate
the whole stress building process in the system. The Lenard-Jones
potential is fixed, so that the repulsive part gives a strong
hard-core repulsion for distances smaller than $2R$, and a small
attractive contribution to the net force between pairs of disks
for distances bigger than $2R$. The radius of disks were
considered as unity ($R = 1$), defining the unit length in the system.
We remain thus with three main parameters governing the generated
patterns: $F_b$, $F_f$ and $\varrho$. The influence of these three
parameters on the final structure was investigated and the
obtained structures were visually compared with experimentally
obtained ones. It is worth mentioning here, that for a successful
simulation the values of $\varrho$ and $F_b$, has to be
correlated. For small $\varrho$ values the value of $F_b$ has to
be chosen high enough, otherwise the majority of springs would
break in the very first simulation step, and no clusterization
would be observable.

The experimental samples were prepared using the drop-coat method
\cite{haynes2001,murray2004}. Of critical importance for
nanosphere ordering is the chemical treatment of the glass
substrate to render the surface hydrophilic and improve its
wettability. This was achieved by etching the substrate in a
solution of sulphuric acid/hydrogen peroxide (3:1) for a period of
3 hours. The substrates were then washed in a copious amount of
deionized water, immersed in a solution of ultra pure
water/ammonium hydroxide/hydrogen peroxide (5:1:1) for 2 hours and
sonicated in an ultrasonic bath for 1 hour. Finally, the
substrates were thoroughly rinsed in ultra pure water and stored
under ultra pure water. Polystyrene nanospheres of 220 nm
diameter, exhibiting negatively charged carboxyl-terminated
surface with a strongly hydrophobic nature, were supplied as
monodispersed suspensions in deionised water (wt 4\%). The
original suspension of polystyrene nanospheres was diluted by 10
and a volume of 100 l diluted solution was evenly spread on the
pre-treated substrates. Finally, samples were dried in an oven at
65 $^0$C for 45 minutes. As the water evaporates the nanospheres
self-assemble into close-packed monolayer arrays exhibiting many
fracture lines, dislocations and other type of defects
(Figure~\ref{fig1}). The microstructure of the sample was studied
by scanning electron microscopy (SEM) using a JEOL JSM 5600 LV
electronic microscope. By surveying the substrate, surface regions
with different morphologies were found even within the same
sample. These qualitatively different structures will arise mainly
due to the non-homogenous nanosphere concentration in different
parts of the sample.

\begin{figure}
 \includegraphics[width=7cm]{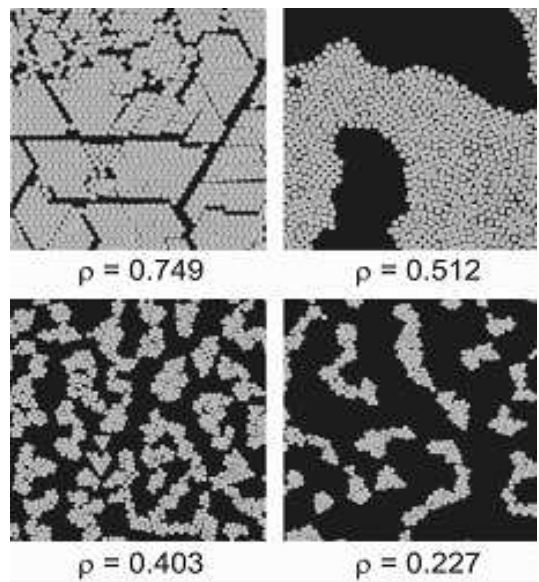}%
 \caption{Characteristic simulation results for the influence of nanosphere
          density on final patterns.\label{fig3}}
 \end{figure}

On Figure~\ref{fig3} we show computer simulation results for the
influence of nanosphere density on the final structure. For high
densities, in agreement with experimental results, we obtain
close-packed domains (triangular lattice structure), separated by
realistic fracture lines and defects. As the nanosphere density
decreases we reproduce the experimentally observed isolated and
elongated islands.

\begin{figure}
 \includegraphics[width=8.5cm]{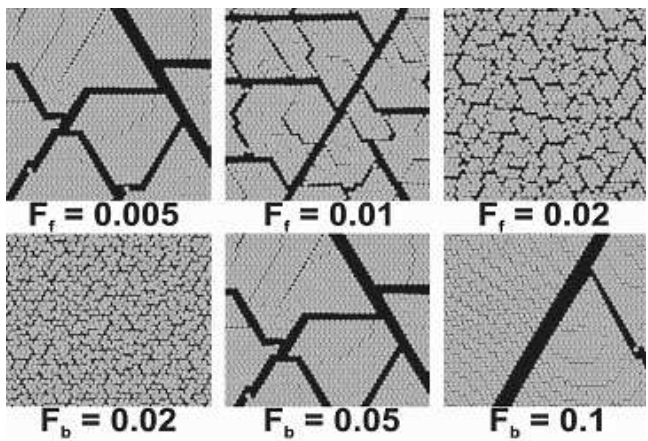}%
 \caption{Simulation results investigating the influence of parameters
          $F_f$ and $F_b$ on the final patterns. On the top line $F_f$ is
          changing ($F_b = 0.05$), while on the bottom line the varied
          parameter is $F_b$ ($F_f = 0.005$). For all these simulations
          the other parameters were chosen as: $k_{ini} = 0.1$, $dk = 0.002$,
          $\varrho = 0.749$, $\eta = 250$.\label{fig4}}
 \end{figure}

Similarly with the results obtained by Leung et al.
\cite{leung2000} we observed that the topology of the final
structure is strongly affected by the values of $F_f$ and $F_b$.
Increasing the value of $F_f$ will result in denser and thinner
fracture line structure with smaller close-packed domains (top
line of Figure~\ref{fig4}). Decreasing the value of $F_b$ results
in a similar effect (bottom line of Figure~\ref{fig4}).
Experimentally one could control $F_f$ by changing the substrate
and $F_b$ by changing the evaporating liquid.

\begin{figure}
 \includegraphics[width=7cm]{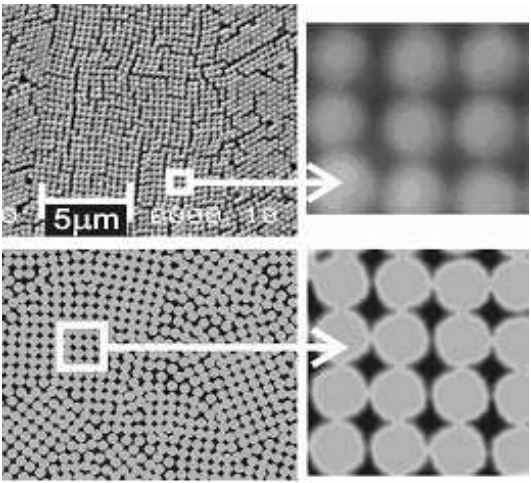}%
 \caption{Experimental (top) and simulation results (bottom) for the square
          lattice crystallization phase.\label{fig5}}
 \end{figure}

The dynamics leading to formation of final structures can be also
easily revealed through simulations (to do that experimentally is
quite a difficult task). Some movies in this sense are presented
on the home-page dedicated to this study \cite{jarai}. The
formation of close packed (triangular lattice) domains separated
by characteristic fracture lines is realized through a series of
stick-slip avalanches, as the value of the spring constant is
increased. The initial nanosphere configuration is quickly lost,
and due to long-range elastic forces an intermediate square
lattice multi-domain structure is formed (Figure~\ref{fig5},
bottom line). This intermediate (metastable) structure is lost as
$k$ increases further (long springs are broken) and slowly the
triangular symmetry is selected. During the relaxation of stress
and non-homogeneity in the system, fracture lines are nucleated
and propagated.  This process will lead to the final
characteristic pattern.

The intermediate square lattice structure can be also revealed by
experiments, either by analyzing samples that are not completely
dried, or by considering a quick, non quasi-static drying
procedure. It is believed \cite{ong} that in case of non
quasi-static drying the nanosphere system is blocked in this
intermediate, metastable phase. Experimentally, we have also
obtained samples that show this intermediate square-lattice
crystallization phase (Figure~\ref{fig5}, top line). The success
in reproducing this square-lattice crystallization phase and in
general the qualitative agreement between simulation results
(Figure~\ref{fig2}-\ref{fig3}) and the experimentally observed
structures (Figure~\ref{fig1}) encourages us to believe that the
introduced model is a good one and the picture we got for the
underlying dynamics is correct.

We have introduced thus a successful model for describing
monolayer patterns that are formed on a surface after quasi-static
drying of a colloidal nanosphere suspension.  By using this model
one can get a first picture on the underlying nano-scale dynamics
and can analyze the influence of the experimentally adjustable
parameters on these practically important structures.

\bibliography{nanospheres}

\end{document}